\documentclass[11pt,letterpaper]{article}

\usepackage{pstricks,fullpage,  epsfig,amsthm,amsmath,latexsym,amssymb,verbatim,url,setspace,multirow,fancyhdr,
pdfsync,epstopdf,rotating,xfrac,footnote,pst-plot}

\usepackage[para,online,flushleft]{threeparttable}

\usepackage[hidelinks]{hyperref}
\hypersetup{
    colorlinks=false,       
    linkcolor=red,          
    citecolor=green,        
    filecolor=magenta,      
    urlcolor=blue           
}

\usepackage{sectsty}

\sectionfont{\centering\fontsize{14}{15}\selectfont}
\subsectionfont{\centering\fontsize{12}{13}\selectfont}

\usepackage{microtype}
\usepackage{rotating}
\usepackage[round,comma]{natbib}
\usepackage[bottom]{footmisc}
\usepackage[font=footnotesize,labelfont=bf]{caption}
\usepackage{subcaption}
\usepackage{mathtools}
\DeclarePairedDelimiter{\ceil}{\lceil}{\rceil}
\DeclarePairedDelimiter{\floor}{\lfloor}{\floor}
\DeclareFontSeriesDefault[rm]{bf}{b}

\usepackage[
    linecolor=red,
    bordercolor=red,
    backgroundcolor=white,
    textsize=tiny
]{todonotes}

\usepackage{marginnote}

\newcommand{\ltodo}[2][]{{%
 \let\marginpar\marginnote
 \reversemarginpar

 \renewcommand{\baselinestretch}{1.5}%
 
 \todo[#1]{#2}}}

\newcommand{\noi}{\noindent}

\newcommand{\st}{~|~}

\newtheorem{lemma}{Lemma}
\newtheorem{theorem}{Theorem}

\newtheorem{proposition}{Proposition}

\newtheorem{remark}{Remark}

\renewenvironment{proof}{\noi \textbf{Proof.}}{$\quad
\blacksquare$\\}

\usepackage[
    linecolor=red,
    bordercolor=red,
    backgroundcolor=white,
    textsize=tiny
]{todonotes}

\newcounter{Lcount}

\graphicspath{{Figures/}}

\title{\textbf{Motivated Reasoning and Information Aggregation}\footnote{We are grateful to Andrew Little, Peter Buisseret, Justin Grimmer, and  audiences at Stanford and the University of Chicago  for helpful comments and/or conversations.}}

\author{\textbf{Avidit Acharya}\footnote{Professor, Stanford Political Science Department (email: avidit@stanford.edu).} \and \textbf{Kyungtae Park}\footnote{PhD student, Stanford Political Science Department (email: pkpark@stanford.edu).} \and \textbf{Tomer Zaidman}\footnote{PhD student, Stanford Political Science Department (email: tzaidman@stanford.edu).}}

\date{December 2025}

\begin{document}

\setstretch{1.05}

\maketitle

\begin{abstract}
If agents engage in motivated reasoning, how does that affect the aggregation of information in society? We study the effects of motivated reasoning in two canonical settings---the Condorcet jury theorem (CJT), and the sequential social learning model (SLM). We define a notion of motivated reasoning that applies to these and a broader class of other settings, and contrast it to other approaches in the literature. We show for the CJT that information aggregates in the large electorate limit even with motivated reasoning. When signal quality differs across states, increasing motivation improves welfare in the state with the more informative signal and worsens it in the other state. In the SLM, motivated reasoning improves information aggregation up to a point; but if agents place too little weight on truth-seeking, this can lead to worse aggregation relative to the fully Bayesian benchmark. \medskip

\end{abstract}

\clearpage

\section{Introduction}

We study how information aggregation in two canonical models---the Condorcet voting model studied by \citet{de2014essai}, \citet{austen1996information}, \citet{feddersen1997voting}, and many others, and the sequential social learning model introduced by \citet{bhw}---is affected when agents are motivated reasoners. Our theory of motivated reasoning builds directly upon the foundational principles of probability theory and the perspective in \citet{acharyaExplainingPreferencesBehavior2018} and \citet{littleHowDistinguishMotivated2025} that motivated reasoning is the outcome of a dissonance reduction problem.

In our applications, all agents receive private information about a binary state of the world, and we specify that agents are motivated to believe that the state (imperfectly) indicated by their private signal is the true state. Given a binary state $\theta \in \{A, B\}$, if an agent with prior $P$ receives a signal $s$ such that $P(\theta \st s) > P(\theta)$ then their motivated belief is a probability measure 
$$\hat P (\cdot \st s) = w P(\cdot\st \theta, s) + (1-w) P(\cdot\st s )$$
where $w \in [0,1]$ is the weight the agent puts on motivation and $1-w$ the weight they put on truth-seeking: $P(\cdot \st s)$ is the probability conditional on signal $s$ and $P(\cdot\st \theta, s)$ is the probability  conditional on both $s$ and state $\theta$ being the true state. Thus when an agent receives a signal $s$ in favor of a state $\theta$, they would like to believe that $\theta$ is in fact the true state, placing weight $w$ on that belief and $1-w$ on the truth. 

Before turning to these applications, we develop a general theory of motivated reasoning as the solution to a cognitive dissonance reduction problem. Our measure of dissonance corresponds to the $w$-weighted sum of the squared $L^2$ distance between the objective belief, which is $P(\cdot~|~s)$ in the example above, and the motivation $P(\cdot~|~\theta, s)$ in that example. This gives the formula for the motivated belief $\hat P(\cdot~|~s)$ stated above, and contrasts with other measures of belief dissonance that are used in the literature, including the influential KL-divergence-based measure used by \citet{littleDistortionRelatedBeliefs2019,littleHowDistinguishMotivated2025}. In the applications, we show that the degree of motivation $w$ affects welfare, and we analyze its effects. 

In the canonical Condorcet voting model there is a binary state $\theta \in \{A, B\}$ corresponding to the guilt and innocence of a given suspect, and all agents receive one of two i.i.d.~signals $s \in \{a, b\}$. Agents vote simultaneously, under majority rule, for one of two alternatives. Suppose the signal accuracy of the confirmatory signal is higher in state $A$ than it is in state $B$, i.e. $q_A = \Pr[s = a~|~\theta = A]  > \Pr[s = b~|~\theta= B] = q_B > 1/2$. In this setting, we find that motivated reasoning moves the equilibrium vote probabilities in the direction of sincere voting, and  that information continues to aggregate in the large election limit   \citep[though this result is not implied in our setting by][]{mclennan1998consequences}. In addition, for large electorates, as the weight on motivation $w$ increases, welfare (i.e.,~the probability of making the correct collective decision) increases when the true state is $A$ but decreases when it is $B$. Finally, we use a recent bound from \citet{ferrante2021bounds} to show that as the number of voters grows to infinity, the convergence rate for welfare is completely independent of the degree of motivation $w$. The reason for this surprising result is that the equilibrium voting strategy converges to its limit much faster than the rate of convergence of welfare. 

Turning to the sequential learning model, recall that in the  canonical setting of \citet{bhw}, there is a binary state $\theta \in \{A,B\}$ indicating which of two restaurants is better, and each agent in an infinite sequence receives one of two possible i.i.d.~signals $s \in \{a,b\}$. The agents sequentially decide which restaurant to go to based on their own private signal and observations of their predecessors' choices. The signals are such that $\Pr[s = a~|~\theta = A] = \Pr[s = b~|~\theta = B] = p > 1/2$. Impose the tie-breaking rule that when indifferent an agent will follow their own signal. As long as the agents are following their own signals, once one restaurant has attracted two more customers than the other, all subsequent agents ignore their signals and herd to that restaurant. The probability that there will eventually be an informational cascade to the better restaurant is $p^2/[p^2+(1-p)^2]$. However, when agents are motivated reasoners who put weight $w$ on motivation, there is an integer $k^*(w) \geq 2$ that is unboundedly increasing in $w$ up to $w = 1/2$, such that the probability of eventually cascading to the better restaurant is $p^{k^*(w)}/[p^{k^*(w)}+(1-p)^{k^*(w)}]$. Thus, welfare is increasing to 1 in $k^*(w)$. When $w \geq 1/2$, then all agents follow their own private signals. In this case, welfare is only $p$, the share of agents that enter the better restaurant. This is lower than welfare in the case of any $w < 1/2$, under which a cascade always eventually occurs. 

Motivated reasoning affects behavior by encouraging agents to follow their own private signals. At moderately low levels, it delays the onset of a cascade and improves the odds of eventually cascading to the better restaurant. The hidden information of a larger set of earlier agents is unlocked for later agents to exploit. However, these earlier agents suffer as many of them make the wrong choice. They do not internalize the information that their predecessors' choices reveal to the same extent that fully Bayesian agents would. We formally analyze this exploration-exploitation trade-off in the case of a finite number of agents and prove that the optimal level of motivation increases with population size. Motivation here directly affects welfare, but as before, does not  affect its rate of convergence. Although \citet{tamuzetal} uncover similar results in a richer informational model in which agents have misspecified beliefs, our analysis provides a sharper understanding of the exploitation-exploration tradeoff (particularly in the finite agent case) by specializing to the case of binary signals. 

\section{Motivated Reasoning}

The concept of motivated reasoning stems from a long line of work in social psychology beginning with \citet{festingerTheoryCognitiveDissonance1957} on cognitive dissonance and \citet{kundaCaseMotivatedReasoning1990} on the tradeoff between truth-seeking and directional motives in belief formation. Here, we offer a model of motivated reasoning as the outcome of a dissonance reduction problem, as suggested by \citet{acharyaExplainingPreferencesBehavior2018} and \citet{littleDistortionRelatedBeliefs2019,littleHowDistinguishMotivated2025}, and others across a variety of disciplines including \citet{harmon-jonesActionBasedModelCognitiveDissonance2015}, and \citet{mcgrathDealingDissonanceReview2017} in psychology, \citet{brunnermeierOptimalExpectations2005}, \citet{brachaAffectiveDecisionMaking2012}, \citet{akerlofEconomicConsequencesCognitive1982}, and \citet{golmanPreferenceBeliefConsonance2016} in economics, and \citet{minozziEndogenousBeliefsModels2013} and \citet{penn2017inequality} in political science. The key idea is that agents experience cognitive discomfort when they hold beliefs (or attitudes/values) that do not accord with their behavior or prior preferences, and therefore construct more consonant beliefs to reduce this discomfort.

Given a measurable space $(\Omega,\mathcal F)$, where $\Omega$ is the sample space and $\mathcal F \subseteq 2^{\Omega}$ is a $\sigma$-algebra, let an agent's \emph{objective belief} be a probability measure $P:\mathcal F \to [0,1]$. This may be either a prior belief or an interim posterior obtained by Bayes' rule after observing some signal that we associate with the agent having learned an event in $\mathcal F$. The triple $(\Omega,\mathcal F,P)$ is the probability space that we operate in. The agent's \emph{motivation} is represented by a second probability measure $P_Z$ on $(\Omega,\mathcal F)$, which we interpret as the belief the agent would like to hold if some \emph{motivating event} $Z \in \mathcal F$ were true. When $P(Z)>0$ and we wish to impose Bayesian coherence, we set $P_Z = P(\,\cdot \mid Z)$. When $P(Z)=0$ or when the motivated belief is not literally a Bayesian posterior, we treat $P_Z$ as a primitive describing the agent's desired belief. Because $P$ and $P_Z$ are probability measures on $(\Omega,\mathcal F)$, there always exists a $\sigma$-finite measure dominating both; we take that to be $Q = P + P_Z$. Identifying any measure $R \ll Q$ with its Radon-Nikodym derivative, $dR/dQ$, the agent's \emph{motivated belief} $\hat P$ is a probability measure over $(\Omega, \mathcal F)$ that solves
\begin{equation}
\min_{\substack{\hat P : \mathcal F \rightarrow [0,1]\\ \hat P \ll Q}}
  w \underbrace{\left\| \hat P -  P_Z\right\|_{L^2(Q)}^2}_\text{motivation}
 + (1-w) \underbrace{\left\|\hat P - P\right\|_{L^2(Q)}^2}_\text{truth-seeking},
\tag{$\ast$}
\end{equation}
for some fixed $w \in [0,1]$, where $L^2(Q)$ is the Hilbert space of square-integrable functions with respect to $Q$ and $\|\cdot\|_{L^2(Q)}$ is the usual $L^2$ norm in this space. Thus an agent who is a motivated reasoner chooses $\hat P$ to minimize a weighted average of the squared $L^2$ distances between their final belief and the truth-seeking belief $P$ and between their final belief and the motivating belief $P_Z$. The parameter $w$ is the weight the agent puts on motivation, and $1-w$ is the weight they put on truth-seeking.

\begin{theorem} \label{thm_one}
The unique solution to $(*)$ is the probability measure
$$\hat P = w P_Z + (1-w) P.$$
\end{theorem}

This result follows from the standard projection theorem, but we provide a direct proof in the appendix for transparency and completeness. The result allows us to interpret $w$ as a parameter capturing the strength of the agent's motivation. If $w = 0$ the agent is fully Bayesian while if $w =1$ he is fully motivated. For $w \in (0,1)$ the agent is partially motivated, partially truth-seeking. In applications, the modeler specifies the motivating event $Z$ and/or the motivation $P_Z$ as part of the assumptions of the model, just as they would specify the agent's prior beliefs, information, and utility over outcomes. 

Our model produces non-Bayesian information processing as a consequence of the influence of the motivating event, and therefore differs from the literature on motivated reasoning as biased information avoidance and processing. For example, \citet{rabinFirstImpressionsMatter1999}  operationalize motivated reasoning as a small probability of misreading a signal that conflicts with one's current belief as one which supports it. Similarly,  \citet{gerberMISPERCEPTIONSPERCEPTUALBIAS1999} and  \citet{druckmanEvidenceMotivatedReasoning2019} model agents as misperceiving or distorting the information structure of disconfirming evidence. 

Our approach also differs from optimal expectations models based on behavioral distortions. For example,   \citet{brachaAffectiveDecisionMaking2012} model motivated reasoning as an interpersonal game between a rational processor whose strategies are actions that directly map to utilities, and an emotional processor that supplies beliefs to the rational processor subject to a distaste for dissonance. Similarly,  \citet{brunnermeierOptimalExpectations2005} model agents that form beliefs with knowledge of what actions they will take on the basis of those beliefs. 

Our model is closely related to the work of \citet{littleDistortionRelatedBeliefs2019,littleHowDistinguishMotivated2025} who develops a theory of motivated reasoning in which agents choose beliefs to minimize a dissonance measure based on KL divergence. We instead use the $L^2$ distance to model dissonance for two reasons. First, when two probability measures like $P$ and $P_Z$ are mutually singular (e.g.,~if $P_Z$ is a point mass on an event $Z$ the agent would like to believe while the objective belief $P$ is continuous), the KL divergences $D_{\mathrm{KL}}(P\|P_Z)$ and $D_{\mathrm{KL}}(P_Z\|P)$ are infinite; so a KL-based dissonance measure would be degenerate in such cases. By contrast, once we fix any dominating measure like $Q$, the squared $L^2 (Q)$ distance between the corresponding densities is always finite. Second, KL divergence is based on maximum-likelihood estimation, making it sensitive to model misspecification and measurement error, as argued by \citet{levine2024method}. Quadratic or minimum-distance criteria such as ours are more robust to such perturbations.

\section{The Condorcet Jury Theorem} 

\subsection{Model}
\label{subsection_cjtmodel}

There are two states of the world $\theta \in \{A, B\}$, equally likely. Each agent $n \in \{1, ... ,2N+1\}$ receives a conditionally independent signal $s_n \in \{a, b\}$, such that 
$$ \forall n, \qquad  \Pr \big[ s_n=a \st \theta = A \big] = q_A \in (
\textstyle{\frac{1}{2}},1) , \qquad \Pr \big[ s_n =b \st \theta = B \big] = q_B \in (
\textstyle{\frac{1}{2}},1) $$
Assume without loss of generality that $q_A \geq q_B$, and let $P$ denote the objective common prior over the state and signal spaces $\{A, B\} \times \{a, b\}^{2N+1}$. After receiving their private signals, all agents update their beliefs and must simultaneously vote for one of two alternatives, also called $A$ and  $B$, and votes are aggregated by majority rule. Let $\rho$ denote the policy that is elected. Each agent $n$'s payoff is 
$$u_n = u (\rho, \theta) = \left\{\begin{array}{ll} 1 & \text{if $\rho = \theta$}\\
0 & \text{if $\rho \neq \theta$} \end{array} \right. $$
Thus, policy $A$ gives a payoff of 1 to all agents when the state is $A$ and 0 when the state is $B$, while policy $B$ gives all a payoff of 1 when the state is $B$ and 0 when it is $A$. 

Let $\hat P_s$ denote any agent's posterior belief after seeing signal $s \in \{a, b\}$, which we assume is possibly motivated. In particular, if an agent receives signal $a$, let his motivating event be the event that $A$ is the true state, while if he receives signal $b$, let it be the event that $B$ is the true state. Thus, the objective updated beliefs for agent $n$ following signals $a$ and $b$ respectively are $P_a = P(\cdot | s = a)$ and $P_b = P(\cdot | s = b)$, the motivations are $P_{a,A} = P(\cdot | s = a, \theta = A)$ and $P_{b,B} = P(\cdot | s = b, \theta = B)$, and motivated beliefs are 
\begin{align*}
\hat  P_a &= w P_{a,A} + (1-w) P_a \\
\hat  P_b &= w P_{b,B} + (1-w) P_b
\end{align*}
where $w \in [0,1]$ is the common weight on motivation shared by all agents. Thus, agents are equally motivated to believe that their own private signal is indicative of the true state.

\subsection{Equilibrium Analysis} 

We examine type-symmetric responsive Nash equilibria---hereafter ``equilibria'' in the rest of this section---in which all agents who receive the same signal use the same strategy, and all agents face a positive probability that their vote is pivotal.\footnote{These are not Bayesian Nash equilibria but once we replace the Bayes updated beliefs with the motivated beliefs, the usual fixed point definition of a Nash equilibrium extends to our setting.} Given parameter $N$ of the model, which governs the total number of agents, we denote a type-symmetric strategy profile with the pair $\sigma_N = (\sigma^A_N, \sigma^B_N)$ which are, respectively, the probability that a voter who receives the $a$ signal votes for $A$ and the probability that a voter who receives the $b$ signal votes for $B$. We define \emph{sincere voting} as the profile $(\sigma^A_N, \sigma^B_N) = (1,1)$. 

Let $N_A$ denote the number of votes from agents other than $n$ in favor of $A$. Given $\hat P_s$, agent $n$ who has private signal $s$ weakly prefers to vote for $A$ if $\hat P_s \big[ N_A = N,~\theta = A \big] \geq \hat P_s \big[ N_A = N,~\theta = B \big]$ and weakly prefers to vote for $B$ if the reverse holds. This follows from the standard calculation that no matter how agents update their beliefs, they care only about how their vote affects outcomes in the event that it is  pivotal.

\begin{proposition}
\label{thm_equilibrium} If $q_A = q_B$ then for all $N$  the unique equilibrium of the voting game is sincere voting. If $q_A > q_B$ then let $$ N^*(w) = \ceil[\Bigg]{ \log \psi(w) \big/ \log \frac{q_A(1-q_A)}{q_B(1-q_B)} } \quad \text{where} \quad \psi(w) = \frac{(1-w)(1-q_B)}{q_A + w(1-q_B)}.$$
Then, for $N < N^*(w)$ the unique equilibrium is sincere voting, while for $N \geq N^*(w)$ it is 
$$ \sigma_N = (\sigma^A_N, \sigma^B_N) =  \left(\frac{q_A - \psi(w)^{1/N} (1-q_B)}{(q_A)^2 - \psi(w)^{1/N} (1-q_B)^2},  1 \right)  $$
\end{proposition}

We now make a few observations about the equilibrium. As $w \rightarrow 1$ sincere voting is the unique equilibrium for all $N$, as in this case $N^*(w) \rightarrow \infty$. In fact, $N^*(w)$ is increasing in $w$, and so is the equilibrium mixing probability $\sigma^A_N$ in Proposition \ref{thm_equilibrium}, as 
\begin{equation} \label{eq:sigAw}
\frac{\partial \sigma^A_N}{\partial w} = \frac{d \sigma^A_N}{d \psi^{1/N}} \frac{1}{N} \psi^{\frac{1}{N} -1} \frac{d \psi}{d w} =  -\frac{q_A (1-q_B)(q_A+q_B-1)}{[(q_A)^2 - \psi^{1/N}(1-q_B)^2]^2} \frac{1}{N} \psi^{\frac{1}{N} -1} \frac{d \psi}{d w} > 0 
\end{equation}
which follows because $d\psi/dw < 0$. Therefore, motivated reasoning pushes equilibrium behavior towards sincere voting, and so the question of whether it improves the rate of information aggregation depends on a comparison of aggregation rates under sincere voting to rates under strategic voting in the fully Bayesian  benchmark of $w = 0$. 

Next, we study how the weight on motivation $w$ affects information aggregation both away from the infinite population limit, and the speed of aggregation in the limit. Following the literature, we define a limit equilibrium $\sigma_\infty = (\sigma^A_\infty, \sigma^B_\infty)$ to be the limit of a sequence of equilibria $\{(\sigma^A_N, \sigma^B_N)\}_{N=1}^\infty$ with each $\sigma_N = (\sigma^A_N, \sigma^B_N)$ being the equilibrium of a game with $2N+1$ voters. We measure welfare in a game with population parameter $N$ as the probability of electing the correct policy, i.e.~the probability $W_N (\sigma_N, \theta,w)$ of electing $\rho = \theta$ when the state is $\theta$ under the equilibrium $\sigma_N$ of Proposition \ref{thm_equilibrium}.\footnote{As the game is common interest, this notion of welfare coincides with the expected sum of utilities,  $$\mathbb E_{\sigma_N} \left[ \frac{1}{2N+1} \sum_{i=1}^{2N+1} u (\rho, \theta) \right] = \mathbb E_{\sigma_N,N} [u(\rho,\theta)] =  P_{\sigma_N,\theta,N} (\{\rho = \theta\}) = W_N (\sigma_N, \theta,w).$$}







\begin{proposition}
\label{thm_condorcet}
    Let $\sigma_N$ be the unique equilibrium and $N^*(w)$ the threshold characterized in Proposition \ref{thm_equilibrium} for the voting game with population parameter $N$.
\begin{enumerate}
\item[(i)] There is a unique limit equilibrium $\sigma_\infty$ and $$\lim_{N\rightarrow \infty} W_N (\sigma_N, \theta,w) = 1\qquad \text{for both $\theta \in \{A,B\}$.}$$
\item[(ii)] If $q_A = q_B$ then for all $N$, $W_N (\sigma_N, \theta,w)$ is constant in $w$ for both $\theta \in \{A, B\}$. If $q_A > q_B$ and $N < N^*(w)$ then $W_N (\sigma_N, \theta,w)$ is constant in $w$ for both $\theta \in \{A, B\}$, while if $N \geq N^*(w)$ then $W_N (\sigma_N, A,w)$ is increasing in $w$ and $W_N (\sigma_N, B,w)$ is decreasing in $w$. 
\end{enumerate}
\end{proposition}

The welfare limit in part (i) of Proposition \ref{thm_condorcet} does not follow from \cite{mclennan1998consequences} due to non-Bayesian belief updating \citep[see][for details]{park2025}.
The upshot of part (ii) of the proposition is that motivation increases the probability of collectively making the correct choice in the state whose signal accuracy is greater but lowers it in the state whose signal accuracy is lower. As motivation moves the equilibrium voting strategies closer to sincere voting, an implication of this result is that under sincere voting, the probability of electing the correct policy in the state with better signal accuracy is higher than the same probability under strategic voting; and the opposite is true in the other state. To our knowledge, this observation had not previously been identified in the literature.

Next, we derive the rate of convergence of welfare in $N$. As is standard, we use $\Theta$-notation for the exact convergence rate.\footnote{Formally, for a sequence $\{x_N\}$, $x_N=\Theta(f(N))$ if $a < \textrm{lim\,inf}_{N\rightarrow\infty}x_N/f(N) \le \textrm{lim\,sup}_{N\rightarrow\infty}x_N/f(N)<b$ for constants $a,\,b \in \mathbb{R}$ such that $ab>0$.} 

\begin{proposition}
\label{thm_cjtrate}
Fixing the weight on motivation $w$, let $\sigma_N$ be the unique equilibrium of the voting game with population parameter $N$ given in Proposition \ref{thm_equilibrium} and $\sigma_\infty$ the unique limit equilibrium. Let $\tilde \sigma^\theta_\infty := q_\theta \sigma_\infty^\theta + (1-q_\theta) (1-\sigma_\infty^{\neg\theta})$, using the standard notation that $\neg\theta= B$ when $\theta =A$ and $\neg\theta = A$ when $\theta = B$. Then, for $\theta \in \{A,\,B\},$
$$1-W_N(\sigma_N, \theta,w) = \Theta\left(\left(4 \tilde{\sigma}_\infty^\theta (1-\tilde{\sigma}_\infty^\theta) \right)^{N} \Big/\sqrt{N}\right)$$
\end{proposition}

The proposition is derived from a two-step approximation. As welfare $W_N(\sigma_N\,,\theta,w)$ is a binomial random variable with parameters that depend on the ex ante equilibrium vote probabilities $\tilde{\sigma}^A_N$ or $\tilde{\sigma}^B_N$, the first step applies a result from \citet{ferrante2021bounds} that develops a uniform bound on the probability that a binomial random variable with parameters $2N+1$ and $\tilde{\sigma}^\theta_N$ is below $N$. The second step further approximates welfare under the finite-population strategy to that under the limit voting strategy.

Interestingly, the weight on motivation $w$ does not affect the overall convergence rate of equilibrium welfare. The intuition is that the equilibrium strategy $\sigma^\theta_N$ converges to its limit $\sigma^\theta_\infty$ much faster than the convergence rate of the probability $1-W_N(\sigma_N, \theta,w)$ of making the incorrect decision, which we show formally in the appendix.

\section{Sequential Social Learning}

\subsection{Model}

We study the effects of motivated reasoning on information cascades in the model of \cite{bhw}, hereafter BHW.\footnote{Other papers in the literature consider unbounded signals \citep{smith2000pathological, rosenberg2019efficiency}, their steady-state generalizations \citep{mossel2020social}, or multidimensional states \citep{kartik2024beyond}, but we focus on the canonical setting in BHW to make our basic points.}

Suppose there are two possible restaurants, $A$ and $B$, and an infinite sequence of agents $n \in \{ 1,2,..., \infty \}$. Each agent has prior belief that the two restaurants are equally likely to be the better restaurant. Each agent $n$ receives a private signal $s_n \in \{a, b\}$ about which restaurant is better. Denote by $\theta \in \{A, B\}$ the better restaurant. Then, 
$$ \forall n, \qquad  \Pr \big[ s_n=a \st \theta = A \big] = \Pr \big[ s_n =b \st \theta = B \big] = p \in (
\textstyle{\frac{1}{2}},1) $$
Thus, signal $a$ (imperfectly) indicates that $A$ is better while $b$ (also imperfectly) indicates that $B$ is better. All agents' signals are independently delivered. Let $P$ denote the common prior belief over the state and signal spaces $\{A, B\} \times \{a,b\}^\infty$.  Agents sequentially decide which restaurant to go to, and each would like to go to the better one. Denote by $\rho_n \in \{A,B\}$ the restaurant choice of agent $n$.  
Then, each agent $n$'s payoff is 
$$u_n = u (\rho_n, \theta) = \left\{\begin{array}{ll} 1 & \text{if $\rho_n = \theta$}\\
0 & \text{if $\rho_n \neq \theta$} \end{array} \right. $$
Thus, each receives a payoff of 1 if they go to the better one and 0 if they go to the worse one. Each agent observes the restaurant choice of every agent prior in the sequence to them before making their decision, and in addition to not observing the private signals of these prior agents they also do not observe their payoffs. 

BHW look at perfect Bayesian equilibria (PBE) and assume that when indifferent, an agent will go to each of the two restaurants with equal probability. We, however, drop the requirement that beliefs are updated by Bayes rule whenever possible, and instead assume that agents are motivated reasoners who update their beliefs in a way that we specify below. We also adopt a different tie-breaking rule: if an agent is indifferent given his updated beliefs, he follows his private signal and goes to the restaurant indicated by that signal. We use the term ``equilibrium'' to refer to equilibrium under this tie-breaking rule. 


Following standard terminology, we are in a \emph{cascade} at agent $n$ if that agent goes to the same restaurant given his information about the choices of his predecessors, regardless of his own signal. Once this is true for some agent it is true for all subsequent agents, who then herd to the same restaurant. The question then is whether that restaurant is the better or worse restaurant. Given our tie-breaking rule, any agent can perfectly infer all of the signals of his predecessors from their restaurant choices so long as we are not already in a cascade. Prior to the onset of a cascade, for an agent $n$ who receives private signal $s \in \{a, b\}$ and can infer from the behavior of his predecessors that there were $n_A$ total signals in favor of $A$ and $n_B = n_A - k$ in favor of $B$, the objective updated belief that $A$ is better is 
\begin{align*}
P \big[\theta = A~\big|~s,\text{``$n_B = n_A-k$''} \big] & =   \frac{p^{n_A+\mathbf{1}_{\{s =a\}}} (1-p)^{n_B+\mathbf{1}_{\{s =b\}}}}{p^{n_A+\mathbf{1}_{\{s =a\}}} (1-p)^{n_B+\mathbf{1}_{\{s =b\}}} + (1-p)^{n_A+\mathbf{1}_{\{s =a\}}} p^{n_B+\mathbf{1}_{\{s =b\}}}} \\ &= \frac{1}{1+\left(\frac{1-p}{p} \right)^{{ k+\mathbf{1}_{\{s =a\}} - \mathbf{1}_{\{s =b\}}}} }
\end{align*}
where ``$n_B = n_A-k$'' is our shorthand for the event that there is a total of $n_A$ inferred signals for $A$ and $n_B = n_A -k$ inferred signals for $B$ among all predecessors. The objective updated belief that $B$ is the better restaurant is $P \big[\theta = B~\big|~ s,\text{``$n_B = n_A-k$''} \big]  = 1- P \big[\theta = A~\big|~ s,\text{``$n_B = n_A-k$''} \big] $. However, we assume that agents are motivated reasoners, and if the agent's private signal is $s = a$, then his motivating event is the event that $A$ is the better restaurant, while if it is $s = b$ then it is the event that $B$ is the better restaurant. Thus, if the agent's private signal is $s = a$ his motivation is $P[ \cdot \st \text{``$n_B = n_A-k$''},~s = a,~\theta = A]$ and if it is $s = b$ then it is $P[ \cdot \st \text{``$n_B = n_A-k$''},~s = b,~\theta = B]$. Therefore, if the agent's private signal is $s$ the motivated belief that $A$ is the better restaurant is 
\begin{align*}
\hat  P_{s,k} \big[\theta = A \big] 
& = w \mathbf{1}_{\{s=a\}}+ (1-w) \frac{1}{1+\left(\frac{1-p}{p} \right)^{{ k +\mathbf{1}_{\{s=a\}}-\mathbf{1}_{\{s=b\}}}}}
\end{align*}
and the motivated belief that $B$ is the better restaurant is $\hat P_{s,k}[\theta = B] = 1- \hat P_{s,k} [\theta = A]$. In what follows, we will derive from these expressions the probability that agents will eventually cascade to the better restaurant when they are motivated reasoners.

\subsection{Equilibrium Analysis}

Suppose that the agent described above has signal $s = a$ and is thus motivated to think that $A$ is better. Then for all $k$ such that 
$$ \hat P_{a,k} \big[\theta = A \big] < \frac{1}{2} $$
the agent will go against his private signal and choose restaurant $B$ while for all $k$ such that this inequality doesn't hold, then the agent will choose $A$. Inserting $\hat P_{a,k} \big[\theta = A \big]$ from above, rearranging, and solving for integer $k$, we find that the critical threshold on $k$ is 
$$ k \le  - k^*(w)  := - \Bigg \lfloor \frac{-\log (1-2w)}{\log p - \log (1-p)} \Bigg\rfloor - 2   $$ 
provided $w < 1/2$.  
When $w =0$, for example, and agents are fully Bayesian (or when $w$ is small and they are close enough to fully Bayesian), $k^*(w)$ must equal 2. $k^*(w)$ is nondecreasing in $w < 1/2$. For $w \geq 1/2$, the agent will always follow his private signal and go to restaurant $A$ for all values of $k$. Therefore, for $w \geq 1/2$ we set $k^*(w) = \infty$, indicating that when the strength of motivation is sufficiently strong, an agent who receives a private $a$ signal will never go to restaurant $B$ for any value of $k$. 



The same calculation for an agent who receives signal $b$ and is thus motivated to think that $B$ is better shows that if $ k \geq k^*(w) $  
then the agent will go to $A$ while for all  $k < k^*(w)$ he will follow his private signal and go to $B$. Therefore, there is stochastic process $\{K_n\}$ governing the evolution of $k$ on the set 
$$\mathcal K(w) := \{ -k^*(w), -k^*(w)+1,...,-1,0,1,..., k^*(w)-1, k^*(w) \}$$
such that in equilibrium if $-k^*(w) < K_n < k^*(w)$ then agent $n$ follows his private signal but if the process reaches $k^*(w)$ at some $n$ before it reaches $-k^*(w)$ then a cascade to $A$ will have started at agent $n$ while if it ever reaches $-k^*(w)$ before it reaches $k^*(w)$ then a cascade to $B$ will have started at that point. Once $K_n =-k^*(w)$ or $k^*(w)$ for some $n$, we have $K_{n'} = K_n$ for all $n' > n$. So long as $K_n$ has not touched the boundary values $-k^*(w)$ or $k^*(w)$, it evolves incrementally moving up by one with probability $p$ and down by one with probability $1-p$ when $A$ is the better restaurant, and the reverse probabilities when $B$ is the better restaurant. Thus, $\{K_n\}$ is random walk on the set $\mathcal K(w)$ with absorbing boundaries $-k^*(w)$ and $k^*(w)$. The following proposition summarizes our characterization of equilibrium behavior based on the process $\{K_n\}$. 

\begin{proposition} \label{thm:eqlmslm}
There is a unique stochastic equilibrium sequence of choices $\bar \rho = \{\rho_1,\rho_2,...\}$. In equilibrium, if $K_n \notin \{-k^*(w), k^*(w)\}$ then agent $n$ follows his private signal, i.e.~goes to restaurant $A$ if $s_n = a$ and $B$ if $s_n = b$; if $K_n = k^*(w)$ then both types of agent $n$ go to restaurant $A$; and if $K_n = -k^*(w)$ then both types of agent $n$ go to $B$. 
\end{proposition}

We now investigate the probability of eventually cascading to the better restaurant. By the characterization of $\{K_n\}$ above, this reduces to the problem of finding the probability that if we repeatedly toss a biased coin that shows heads with probability $p > 1/2$, we will eventually observe $k^*(w)$ more heads than tails before we observe $k^*(w)$ more tails than heads---a textbook gambler's ruin problem; see, e.g. \citet{feller1950introduction}. 
We report the solution below and provide a proof in the appendix for completeness.

\begin{proposition} \label{thm_infprob}
If the strength of motivated reasoning is $w \geq 1/2$ then in equilibrium no cascades ever occur in equilibrium. If it is $w < 1/2$ then in equilibrium a cascade occurs almost surely, and agents cascade to the better restaurant with probability 
$$ \phi_{k^*(w)} = \frac{p^{k^*(w)}}{p^{k^*(w)} + (1-p)^{k^*(w)}} $$ 
\end{proposition}

For $k^*(w) = 2$ we have agents who are fully Bayesian and in this case, the probability of cascading to the better restaurant coincides with what has been known in the case of fully Bayesian agents. The probability of not eventually being in a cascade to either restaurant is zero, and so the probability of eventually being in a cascade to the worse restaurant is $1-\phi_{k^*(w)}$. Inspecting these probabilities, it is clear that the probability of eventually cascading to the better restaurant is increasing in $k^*(w)$, which in turn is increasing in the strength of motivation $w$ for $w < 1/2$.

\begin{figure}
\begin{center}
\includegraphics[scale=0.35]{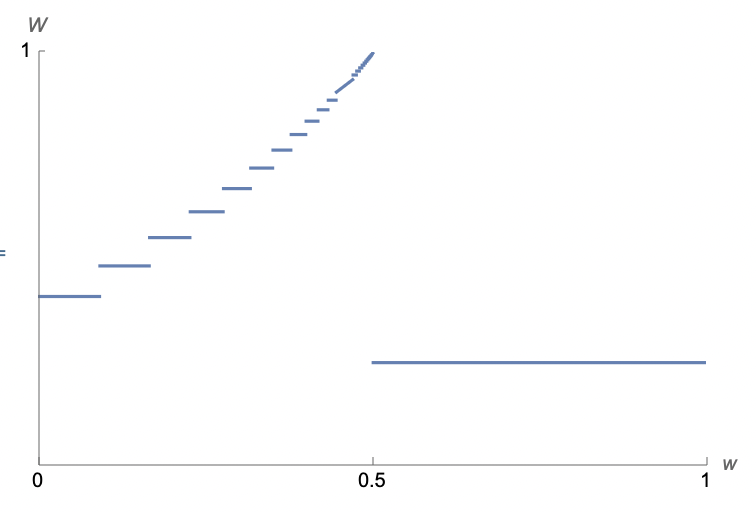}
\end{center}
\caption{Equilibrium welfare in the SLM as a function of the weight on motivation, $w$. \label{fig:welfare}}
\end{figure}

To measure welfare, we take the limiting expected average sum of utilities of the agents; that is, if $\bar \rho = \{\rho_1,\rho_2,...\}$ is the stochastic equilibrium sequence of restaurant choices and fixing $w$ while conditioning on $\theta$ being the better restaurant, welfare is 
$$ W(\bar \rho, \theta,w) =  \lim_{n \rightarrow \infty} \mathbb E \left[ \frac{1}{n} \sum_{i=1}^n u (\rho_i, \theta) \right] $$ 
Given the equilibrium of Proposition \ref{thm:eqlmslm} and the fact that the model is symmetric, this ex ante notion of welfare is independent of which of the two restaurants is better; thus we may write $W(\bar \rho, A, w) = W(\bar \rho, B, w) = W(\bar \rho,w)$. As we show in the course of proving Proposition \ref{thm_infprob} that a cascade initiates almost surely following a finite number of agents when $w<1/2$, welfare in this case is simply the probability that there will eventually be a cascade to the better restaurant, $W(\bar \rho, \theta,w) = \phi_{k^*(w)}$.  When $w \geq 1/2$ and thus $k^*(w) = \infty$, welfare is the limiting share of agents who go to the correct restaurant, which by the law of large numbers is just $p$. The next result summarizes and Figure \ref{fig:welfare} illustrates.

\begin{proposition}
On the range $w \in [0,1/2)$ equilibrium welfare is $W(\bar \rho,w) = \phi_{k^*(w)}$, which is increasing in the strength of motivated reasoning $w$. Then at $w = 1/2$ equilibrium welfare discontinuously drops to $W(\bar \rho,w) = p$ on the range $w \in [1/2,1]$. 
\label{thm_infwelfare}
\end{proposition}

Since $p < \phi_{k^*(w)}$ for all $2 \leq k^*(w)< \infty$, we conclude that motivated reasoning enhances information aggregation up to a point. Too much motivated reasoning to the point where all agents always act only on their private information is harmful for information aggregation leading to worse welfare than in the $w = 0$ case, where agents are fully Bayesian. 

The logic behind Proposition \ref{thm_infwelfare} is as follows. For $k^*(w)$ large but finite, a large but almost surely finite number of agents will go to the restaurant indicated by their private signals. Many of these early agents will make the wrong restaurant choice; but the more they are (which happens when $w < 1/2$ is higher), the more accurate will be the information (about which restaurant is actually better) that their behavior reveals to the infinite number of agents that come after them. Because the welfare of these infinite  agents overwhelms that of the expected finite number of early agents who follow their private signals, increasing the weight on motivation unambiguously improves aggregate welfare so long as cascades eventually happen---and this improvement occurs despite these later agents also being motivated reasoners. Only when the weight on motivation is so high that cascades never happen does aggregate welfare suffer due to motivated reasoning.



\subsection{Finite Population Analysis}

The discussion of welfare in the preceding section revealed a potential tradeoff between exploration and exploitation, but this tradeoff is obscured in the infinite agent case: when $k^*(w) < \infty$, the social cost to a finite number of early agents following their private signals is always more than compensated by the information this reveals to the infinite number of later agents positioned to exploit this information. The same tradeoff is starker in the finite agent case: by the time a cascade would have occurred we may have run out of agents, so delaying the onset of a cascade is not unambiguously good for welfare. Therefore, to study this tradeoff, we assume here that there is a finite number of agents $N$ and ask how the optimal degree of motivation $w$ varies with $N$.

Given the structure of the model (all agents make a single decision sequentially) it is clear that the equilibrium analysis of the infinite agent case carries over to the finite agent case: for any $N$ the unique stochastic equilibrium sequence of restaurant choices $ \bar \rho_{N} =\{\rho_1,...,\rho_N\}$ is simply the first $N$ elements of the stochastic equilibrium sequence $\rho$ of the infinite agent model. We then define welfare analogously. Fixing $\bar \rho_{N}$, equilibrium welfare is 
$$ W_N(\bar \rho_{N}, \theta,w) =\mathbb{E}\bigg[\frac{1}{N} \sum_{n=1}^N u(\rho_n,\,\theta)\bigg].$$
As in the infinite agent model, equilibrium welfare is independent of the true state so we may write $W_N(\bar \rho_{N}, A, w) = W_N(\bar \rho_{N}, B, w) = W_N(\bar \rho_{N}, w)$. Next, for all $N \geq 4$ let 
$$ 
w^*(\bar \rho_{N}, N) = \arg \max_w W_N(\bar \rho_{N}, w) 
$$
be the set of welfare-optimal weights that agents put on motivation. In all results that follow, we fix $\bar \rho_N$ as the stochastic equilibrium sequence of the $N$ agent model, and $\bar \rho$ as the stochastic equilibrium sequence of the infinite agent model. 
\begin{proposition}~
\label{thm:finiteagent}
\begin{enumerate}
\item[(i)] 
Fix $w < \frac{1}{2}$. Then, $W_N(\bar \rho_N, w)$ increases in $N$, converging to $W(\bar \rho, w)$ in the limit. 
\item[(ii)] For all $N \geq 4$,  $\sup w^*(\bar \rho_N, N) < \frac{1}{2}$ is nondecreasing in $N$, and $\lim_N \inf w^* (\bar \rho_N, N) = \frac{1}{2}$.
\end{enumerate}
\end{proposition} 

The fact that $W_N(\bar \rho_N, w)$ increases in $N$, as reported in part (i) of Proposition \ref{thm:finiteagent}, supports the intuition that a larger $N$ cannot harm welfare as a larger population benefits from the information produced by any length of exploration. Part (ii) of the proposition formalizes the tradeoff between exploration and exploitation that motivated our analysis of the finite agent case. With a finite number of agents, the optimal level of motivation $w^*(\bar \rho_N, N)$ is bounded strictly below $1/2$. In addition, as $w^*(\bar \rho_N, N)$ is nondecreasing in $N$, increasing the number of agents increases the benefits of longer exploration. Finally, as the number of agents grows, the optimal weight on  motivation converges to $1/2$.  

Finally, we study the rates at which cascades occur and welfare converges, as we did for the Condorcet voting model. To state the result, let $n^*$ be a random variable indicating the agent number at which a cascade occurs when the threshold is $k^*(w)$. 

\begin{proposition}
\label{thm_social_rate}
    Fix $w < 1/2$. Then,  
\begin{enumerate}
    \item[(i)] $1-\Pr[n^*\le N] =\Theta \big([4p(1-p) \cdot \cos^2 (\pi/2k^*(w))]^{N/2} \big)$
    \item[(ii)] $W(\bar \rho,w) - W_N(\bar \rho_N, w)  = \Theta(1/N)$
\end{enumerate} 
\end{proposition}

Proposition \ref{thm_social_rate} shows that the cascade occurs much faster than welfare converges. This is because the benefit or penalty resulting from agents cascading to either choice permanently pulls welfare in either direction. 

Our results contrast with those of \citet{tamuzetal}, who study a misspecified social learning model in the \citet{smith2000pathological} framework with unbounded signals. In their setting, agents are fully Bayesian but operate within a misspecified model of others’ information quality, and because signals are unbounded, herding is never permanent and society eventually learns the true state. Their central question concerns how the speed of learning varies with the degree and nature of the misspecification, with learning rates governed by the tail behavior of posterior beliefs, since only extreme signals affect outcomes in their continuous signal environment. By contrast, in our binary-signal model, cascades are permanent, herding can occur to the worse restaurant, and motivated reasoning drives departures from Bayesian updating. Moreover, unlike \citet{tamuzetal}, our analysis delivers sharp characterizations of  equilibrium behavior and welfare, and highlights a clear exploration–exploitation tradeoff that is tractably analyzed in the finite agent case. 

\section{Conclusion}

We have provided a general theory of motivated reasoning as the solution to a cognitive dissonance reduction problem and applied it to two canonical settings of information aggregation. 
Because our formulation of motivated reasoning applies to any environment in which agents update beliefs, it can be applied to other settings where agents are asymmetrically informed. It would be particularly interesting to investigate how motivated reasoning shapes interactions in principal–agent and contracting problems, in cheap talk and other communication settings, and in environments of information design, where the interaction between motivated receivers and strategic signal designers remains largely unexplored. These extensions may reveal new sources of inefficiency, comparative statics linking incentives and belief distortions, and conditions under which motivated reasoning can either facilitate or hinder information transmission. These are fruitful avenues for future research.

\clearpage

\appendix

\setstretch{1.00}
\section*{Appendix} \bigskip

\section{Results from Section 2}

\subsection{Proof of Theorem \ref{thm_one}}

As we identify probability measures with their densities in $L^2(Q)$, rewrite problem \textup{($\ast$)} more explicitly as
\[
\min_{h \in \mathcal H}
  w \left\| h - \frac{dP_Z}{dQ}\right\|_{L^2(Q)}^2
 + (1-w) \left\|h- \frac{dP}{dQ}\right\|_{L^2(Q)}^2
\]
where
\[
\mathcal H \;:=\; \left\{ h \in L^2(Q) : h \ge 0 \ Q\text{-a.e.}, \ \int h \, dQ = 1 \right\}
\]
is the set of densities of probability measures that are absolutely continuous with respect to $Q$. For any $h \in L^2(Q)$, expanding the squared norms and collecting terms yields
\begin{align*}
w \left\| h - \frac{dP_Z}{dQ}\right\|_{L^2(Q)}^2
  + (1-w) \left\|h- \frac{dP}{dQ}\right\|_{L^2(Q)}^2 
= \left\| h - \Big[w \frac{dP_Z}{dQ} + (1-w)\frac{dP}{dQ}\Big]\right\|_{L^2(Q)}^2 &  \\[1em]
   + \text{ terms that do not involve $h$ } & 
\end{align*}
which follows after taking inner products and rearranging. Therefore, the unconstrained objective is uniquely minimized at
\[
h = h^\ast := w \frac{dP_Z}{dQ} + (1-w)\frac{dP}{dQ}
\]
in $L^2(Q)$. It is clear that $h^\ast \in \mathcal H$. 

Finally, define $\hat P^\ast$ to be the probability measure with density $h^\ast$ with respect to $Q$. For any event $A \in \mathcal F$, we have
\begin{align*}
\hat P^\ast(A)
& = \int_A h^\ast\, dQ\\
& = w \int_A \frac{dP_Z}{dQ} dQ + (1-w)\int_A \frac{dP}{dQ} dQ  \\
& = w P_Z(A) + (1-w) P(A).
\end{align*}
Therefore, $$\hat P^\ast = w P_Z + (1-w)P,$$ which proves Theorem~\ref{thm_one}.

\section{Results from Section 3}

\subsection{Proof of Proposition \ref{thm_equilibrium}}

For agents with motivated beliefs $\hat P_a$ and $\hat P_b$, the motivated probabilities of the state being $\theta = \tilde \theta \in \{A,B\}$ and the pivotal event occurring are: 
\begin{align*}
\hat P_a (N_A  = N,~\theta = \tilde \theta) &= (1-w) P_a(N_A =N,\,\theta =\tilde\theta) + w P_{a,\,A}(N_A =N,\,\theta =\tilde\theta) \\
&=  w\varphi_A \mathbf{1}_{\{\tilde\theta = A\}} + (1-w) \varphi_{\tilde \theta}   \frac{q_A \mathbf{1}_{\{\tilde\theta=A\}} + (1-q_B) \mathbf{1}_{\{\tilde\theta=B\}}}{q_A + 1-q_B}\\
\hat P_b (N_A  = N,~\theta = \tilde \theta) &= (1-w) P_b(N_A =N,\,\theta =\tilde\theta) + w P_{b,\,B}(N_A =N,\,\theta =\tilde\theta) \\
&=  w\varphi_B  \mathbf{1}_{\{\tilde\theta = B\}} + (1-w) \varphi_{\tilde \theta}   \frac{q_B \mathbf{1}_{\{\tilde\theta=B\}} + (1-q_A) \mathbf{1}_{\{\tilde\theta=A\}}}{q_B + 1-q_A}
\end{align*}
where 
\begin{align*}
\varphi_{A} &= \binom{2N}{N} [ q_{A} \sigma^A_N + (1-q_A) (1-\sigma^B_N) ]^N [  q_{A} (1-\sigma^A_N) + (1-q_{A}) \sigma^B_N ]^N \\
\varphi_{B} &= \binom{2N}{N} [ q_{B} \sigma^B_N + (1-q_B) (1-\sigma^A_N) ]^N [  q_{B} (1-\sigma^B_N) + (1-q_{B}) \sigma^A_N ]^N 
\end{align*}
are the pivotal probabilities conditional on the states being $A$ and $B$ respectively. 

Upon signal $a$, voters weakly prefer to vote for $A$ when $\hat P_a (N_A  = N,~\theta = A) \ge \hat P_a (N_A  = N,~\theta = B)$, or
\begin{align*}
w\varphi_A + (1-w) \varphi_{A}   \frac{q_A}{q_A + 1-q_B} \ge (1-w) \varphi_{B}   \frac{1-q_B}{q_A + 1-q_B} 
\end{align*}
which is equivalent to
\begin{align*} 
\frac{\varphi_A}{\varphi_B} \ge \frac{(1-w)(1-q_B)}{q_A + w(1-q_B)} \equiv \underline{t}.
\end{align*}
They weakly prefer to vote for $B$ when  $\varphi_A/\varphi_B \leq \underline{t}$.  Similarly, upon signal $b$, voters weakly prefer to vote for $B$ when $\hat P_b (N_A  = N,~\theta = B) \ge \hat P_b (N_A  = N,~\theta = A)$, or
\begin{align*}
     \frac{\varphi_A}{\varphi_B} \le \frac{q_B + w(1-q_A)}{(1-w)(1-q_A)} \equiv \bar{t}.
\end{align*}
and weakly prefer to vote for $A$ when ${\varphi_A}/{\varphi_B} \geq \bar{t}$.  Straightforward algebra shows that $q_A \ge q_B > 1/2$ implies that $\underline{t}<1< \bar{t}$. This implies that voters always vote for $B$ if $\varphi_A/\varphi_B<\underline{t}$ and always vote for $A$ if $\varphi_A/\varphi_B>\bar{t}$. These are not responsive equilibria. Voters vote sincerely if $\underline{t} < \varphi_A/\varphi_B<\bar{t}$, and may mix if $\varphi_A/\varphi_B \in \{\underline{t},\,\bar{t}\}$ for at most one of the signals. Therefore, either $\sigma_A^N =1$ or $\sigma_B^N =1$ in the equilibria of interest.

Suppose $\sigma_A^N = 1$. Then, $q_A\ge q_B$ implies
\begin{align*}
    \frac{\varphi_A}{\varphi_B} = \frac{[q_A + (1-q_A)(1-\sigma_N^B)]^N [(1-q_A)\sigma_N^B]^N}{[q_B \sigma_N^B]^N [q_B(1-\sigma_N^B)+ 1-q_B]^N}< 1 < \bar{t}
\end{align*}
and immediately $\sigma_B^N=1$. Now suppose $\sigma_B^N = 1$ instead of $\sigma_A^N =1$. Then,
\begin{align*}
    \frac{\varphi_A}{\varphi_B} = \frac{[q_A\sigma_N^A]^N [q_A(1-\sigma_N^A) + 1-q_A]^N}{[q_B+(1-q_B)(1-\sigma_N^A)]^N [(1-q_B)\sigma_N^A]^N} \ge \underline{t}
\end{align*}
solves to 
$$\sigma_N^A \le \frac{q_A - \psi(w)^{1/N}(1-q_B)}{q_A^2 - \psi(w)^{1/N} (1-q_B)^2} \equiv \sigma_N^*,$$
defining $\psi(w)=\underline{t}$ to emphasize its dependence on $w$. If $\sigma_N^A \ne \sigma_N^*$, then mixing cannot happen since $\varphi_A/\varphi_B$ is away from the threshold, and in a responsive equilibrium, we need $\sigma_N^A = 1$. If $\sigma_N^A=\sigma_N^*$, then $\sigma_N^*$ must be between 0 and 1, which is when $N \ge N^*(w)$. Therefore,  $(\sigma_N^A,\,\sigma_N^B)=(\sigma_N^*,\,1)$ for $N \geq N^*(w)$, and $(\sigma_N^A,\,\sigma_N^B) = (1,\,1)$ for  $N<N^*(w)$. 

Finally, we need to verify that $\varphi_A/\varphi_B < \bar{t}$ under the equilibrium to justify $\sigma_N^B = 1$. We have shown that $\varphi_A/\varphi_B <\bar{t}$ if $(\sigma_N^A,\,\sigma_N^B)=(1,\,1)$. If $\sigma_N^A = \sigma_N^*$, then $\varphi_A/\varphi_B = \underline{t} < \bar{t}$. 

\subsection{Proof of Proposition \ref{thm_condorcet}}

(i) Suppose the state is $\theta = A$. For all voters $n$ define the random variable $x_n = +1$ if $n$ votes $A$ and $x_n = -1$ if $n$ votes $B$. Consider the random variable $X_N = \sum_{n=1}^{2N+1} x_n$. The ex ante probability of drawing a vote for $A$ is
$$\tilde \sigma^A_N := q_A \sigma_N^A + (1-q_A) (1-\sigma_N^B) \rightarrow \tilde{\sigma}_\infty^A :=\frac{q_A}{1+q_A - q_B} > \frac{1}{2}$$
where $\sigma_N^A$ and $\sigma_N^B$ are the equilibrium vote probabilities from Proposition \ref{thm_equilibrium}, the limit follows because $\psi(w)^{1/N} \rightarrow 1$ as $N \rightarrow \infty$, and the inequality follows by our assumption that $q_A \geq q_B > 1/2$. Set $\eta_N = \tilde \sigma^A_N - 1/2$ for all $N$, and note that $\mathbb E[x_i] = 2 \eta_N > 0$. Then since $\eta_N$ converges to some $\eta_\infty$ there is a sequence $\delta_N \rightarrow 0$ such that
\begin{align*}
W_N(\sigma_N, A) + \delta_N  = \Pr(X_N > 0) + \delta_N  &\geq \Pr \left( \bigg| \frac{X_N}{2N+1} - 2 \eta_\infty \bigg| < 2\eta_\infty \right) + \delta_N  \\
& \geq \Pr\left( \bigg| \frac{X_N}{2N+1} - 2 \eta_N \bigg| < 2\eta_\infty\right)\\ & = 1 - \Pr\left(\bigg| \frac{X_N}{2N+1} - 2 \eta_N \bigg| \geq 2\eta_\infty \right) \\
& \geq 1 - \frac{4\tilde\sigma_N^A (1-\tilde\sigma_N^A)}{(2N+1)4\eta_\infty^2}  \rightarrow 1,
\end{align*}
where the second inequality follows from the definition of $\{\delta_N\}$ and that each probability mass of $X_N/(2N+1)$ uniformly converges to zero, and the third from Chebyshev's inequality. Since $4\tilde\sigma_N^A (1-\tilde\sigma_N^A) \rightarrow 4\tilde\sigma_\infty^A (1-\tilde\sigma_\infty^A)$, the limit in the final line holds. Thus, $\Pr(X_N > 0) \rightarrow 1$ in $N$, and  $W_N(\sigma_N, A,w) \rightarrow 1$. The $\theta = B$ case is similar and omitted. \bigskip

\noindent (ii)
Define $\tilde \sigma^B_N = q_B \sigma_N^B + (1-q_B) (1-\sigma_N^A) = 1-(1-q_B) \sigma_N^A$ analogously to $\tilde{\sigma}_N^A$ above, where $\sigma^A_N$ and $\sigma^B_N$ are the equilibrium vote probabilities for types receiving signals $a$ and $b$ respectively. From the remark following Proposition \ref{thm_equilibrium}, $\tilde \sigma^\theta_N$ increases in $w$ in state $A$ but decreases in it in state $B$. Define a random variable $Z_p$ that follows $\textrm{Binom}(2N+1,\,p)$ for a fixed $N$ and any $p \in [0,\,1]$. Then,
$$ W_N (\sigma_N, \theta,w) = \Pr \big[Z_{\tilde{\sigma}_N^\theta}>N\big].$$
The claim holds if $\Pr\big[Z_p>N\big]$ increases in $p$, or $\Pr\big[Z_p>N\big]>\Pr\big[Z_q>N\big]$ for all $p>q$. Write out $Z_p$ and $Z_q$ as sums of $2N+1$ i.i.d.~Bernoulli random variables $\{X_i\}_{i=1}^{2N+1}$ each with success probability $p$ and $\{Y_i\}_{i=1}^{2N+1}$ each with success probability $q$. Thus, $Z_p = \sum_{i=1}^{2N+1} X_i$ and $Z_q = \sum_{i=1}^{2N+1}Y_i$. We couple the distributions of $X_i$ and $Y_i$ so that $$(X_i,\,Y_i )= \begin{cases}(1,\,1)  & \text{with probability $q$} \\ (1,\,0) & \text{with probability $p-q$} \\ (0,\,0) & \text{with probability $1-p$}\end{cases}$$
independent across $i$. Since $X_i \ge Y_i$ for all $i$, the inequality $\sum_{i=1}^{2N+1} X_i \ge \sum_{i=1}^{2N+1} Y_i$ deterministically holds. Also, $\Pr\big[\sum_{i=1}^{2N+1} X_i >N,\,\sum_{i=1}^{2N+1}Y_i <N\big]>0$. Therefore,
$$\Pr\bigg[ \sum_{i=1}^{2N+1} X_i >N \bigg] > \Pr\bigg[\sum_{i=1}^{2N+1} Y_i>N \bigg] \,\,\, \Longrightarrow \,\,\,\Pr\big[Z_p>N\big]>\Pr\big[Z_q>N\big].$$

\subsection{Proof of Proposition \ref{thm_cjtrate}}
Suppose the state is $\theta = A$. $1-W_N(\sigma_N, A,w)$ is the probability that $B$ is elected. We use the following result from \citet{ferrante2021bounds}: for $X_n \sim \textrm{Binom}(n,\,p)$,
$$\Pr[X_n \le \alpha n] \in  \frac{1}{\sqrt{2\pi \alpha(1-\alpha) n}} e^{-n D_{\mathrm{KL}}(\alpha\,\lVert \,p)} \cdot \bigg[\frac{1-c_\alpha(1/r)/n}{1-1/r},\,\frac{1}{1-r}\bigg]$$
where $r=p(1-\alpha)/\alpha(1-p)$, $c_\alpha(r) = 1/\alpha(1-\alpha) \cdot [1+r(1+r)/(1-r)^2]$, and $D_{\mathrm{KL}}(\alpha\,\lVert \,p)$ denotes the KL-divergence between two Bernoulli distributions $\textrm{Bern}(\alpha)$ and $\textrm{Bern}(p)$. In our setting, $X_N \sim \mathrm{Binom}(2N+1,\,\tilde{\sigma}_N^A)$, $\alpha=1/2$, $r_n = \tilde{\sigma}_N^A/(1-\tilde{\sigma}_N^A)$, $c_\alpha(r_n) =4 + 4r_n(1+r_n)/(1-r_n)^2$. Note that $\tilde{\sigma}_n^A$ converges to a limit strictly bounded away from 1. Therefore, given $n \rightarrow \infty$, both the lower and upper bounds, $\frac{1-c_\alpha(1/r_n)/n}{1-1/r_n}$ and $\frac{1}{1-r_n}$, are $\Theta(1)$. This shows that
\begin{align*}
\Pr\big[X_n\le N\big] & = \Theta\bigg( \frac{1}{\sqrt{2N+1}}e^{-(2N+1) D_{\mathrm{KL}}(1/2\,\lVert \,\tilde{\sigma}_N^A)}\bigg) 
= \Theta\left(\left(4 \tilde{\sigma}_N^A (1-\tilde{\sigma}_N^A) \right)^{N} \Big/\sqrt{N}\right).
\end{align*}
We next show that the convergence rate can be written in terms of $\tilde{\sigma}_\infty^A$ instead of $\tilde{\sigma}_N^A$. Note that $\{\tilde{\sigma}_N^A\}_N$ is a decreasing sequence. The ratio between the true convergence rate and the one replaced with $N=\infty$ is 
\begin{align*}
\left( \frac{\tilde{\sigma}_\infty^A (1-\tilde{\sigma}_\infty^A)}{\tilde{\sigma}_N^A (1-\tilde{\sigma}_N^A)}\right)^{N}   = \left( 1+t_N \right)^{N} =: T_N
\end{align*}
for $t_N  = \big[ \tilde{\sigma}_\infty^A (1-\tilde{\sigma}_\infty^A) - \tilde{\sigma}_N^A (1-\tilde{\sigma}_N^A)\big]/\tilde{\sigma}_N^A (1-\tilde{\sigma}_N^A) \rightarrow 0$. Since $T_N \ge 1$, $T_N = \Theta(1)$ if $t_N = O(1/N)$. Then note that the denominator of $t_N$ converges to a non-zero limit and its numerator is $(\tilde{\sigma}_N^A  - \tilde{\sigma}_\infty^A) (\tilde{\sigma}_N^A  + \tilde{\sigma}_\infty^A -1) = O(\tilde{\sigma}_N^A  - \tilde{\sigma}_\infty^A)$ since $\tilde\sigma_N^A$ is bounded. For large enough $N$,
\begin{align*}
    \tilde{\sigma}_N^A  - \tilde{\sigma}_\infty^A &=  q_A \bigg(\frac{q_A - \psi(w)^{1/N} (1-q_B)}{(q_A)^2 - \psi(w)^{1/N} (1-q_B)^2} -  \frac{1}{1+q_A - q_B}\bigg)\\
    &= O\big(\big[q_A - \psi(w)^{1/N} (1-q_B) \big] \big[1+q_A - q_B\big] -  \big[(q_A)^2 - \psi(w)^{1/N} (1-q_B)^2\big]\big)\\
    &=O\big(1-\psi(w)^{1/N}\big)
\end{align*}
where the second equation is by suppressing the denominator converging to a nonzero limit and the last equation is by factoring out $q_A(1-q_B)$. Since $\psi(w)\in (0,\,1)$ and $\lim_{x \rightarrow 0} (1-e^{-x})/x = 1$, we have $\tilde{\sigma}_N^A - \tilde\sigma_\infty^A = O(1/N)$ and $T_N = O(1)$ and the claim of the proposition follows. An analogous argument can be made for state $\theta =B$.

\section{Results from Section 4}

\subsection{Proof of Proposition \ref{thm_infprob}}

When $w \geq 1/2$ then $k^*(w) = \infty$ while if $w < 1/2$ then $2 \le k^*(w) < \infty$. If $k^*(w) = \infty$, then all agents follow their own signals. 

For the $w < 1/2$ case, and assume without loss that $\theta = A$ be the better restaurant (a similar argument will apply and give us analogous solutions if $\theta = B$). Take the random walk $\{K_n\}$ described in the main text, and let   
$$ n^* = \inf \{ n \geq 0~:~K_n =+k^*(w) \text{ or } K_n = -k^*(w)\}$$
be the time at which one of the two boundaries is hit. Since $\{K_n\}$ is a finite state Markov chain and all states besides $\pm k^*(w)$ are transient, $n^*$ is almost surely finite. Letting 
$\mathcal A$ denote the event $\{K_{n^*} = k^*(w)\}$ and $\mathcal B$ be the event $\{K_{n^*} = -k^*(w)\}$, we therefore have
\begin{equation} \label{prAprB}
    \Pr(\mathcal A) + \Pr (\mathcal B) = 1.
\end{equation}
Next, consider the process 
$$Z_n = \left(\frac{1-p}{p} \right)^{K_n}$$
with $\mathcal F_n$ being the filtration up to agent $n$. Then $(Z_n, \mathcal F_n)$ is a martingale, as 
\begin{align*}
\mathbb E[Z_{n+1}|\mathcal F_n] &= \mathbb E \left[\left( \frac{1-p}{p} \right)^{K_{n+1}}~|~\mathcal F_n \right]  \\
& = \left(\frac{1-p}{p} \right)^{K_n} \left[p \left(\frac{1-p}{p}\right)^{+1} + (1-p) \left(\frac{1-p}{p}\right)^{-1} \right] \\
& = \left(\frac{1-p}{p} \right)^{K_n} \\
& = Z_n.
\end{align*} 
Therefore, the optional stopping theorem implies 
\begin{equation} \label{eq:ost}
1 = \left(\frac{1-p}{p}\right)^0 = \mathbb E [Z_0] = \mathbb E [Z_{n^*}] = \Pr (\mathcal A) \left( \frac{1-p}{p} \right)^{k^*(w)} + \Pr (\mathcal B) \left( \frac{1-p}{p} \right)^{-k^*(w)}
\end{equation}
since at event $\mathcal A$ we have $Z_{n^*} = [(1-p)/p]^{k^*(w)}$ while at $\mathcal B$ we have $Z_{n^*} = [(1-p)/p]^{-k^*(w)}$. Equations \eqref{prAprB} and \eqref{eq:ost} together are two equations in two unknowns, with solutions 
$$ \Pr (\mathcal A) = \frac{p^{k^*(w)}}{p^{k^*(w)} + (1-p)^{k^*(w)}} \quad \text{and} \quad \Pr (\mathcal B) = \frac{(1-p)^{k^*(w)}}{p^{k^*(w)} + (1-p)^{k^*(w)}}. $$

\subsection{Proof of Proposition  \ref{thm:finiteagent}}

Continue to assume that $\theta=A$ is the better restaurant. Recall that $n^*$ is the random variable that indicates the number of periods passed until a cascade occurs when the threshold is $k^* = k^*(w)$. The total utility is
\begin{align*}
\sum_{n=1}^N u(\rho_n,\,\theta) := \sum_{n=1}^{N} \big(A_{n} + (N-n)  \cdot \mathrm{Bern}\big(\phi_{k^*} \big) \big) \cdot \mathbf{1}_{\{\min(n^*,\,N)=n\}} 
\end{align*}
where $A_n$ is the number of ``correct'' choices prior to the start of a cascade (i.e.~choices to go to restaurant $A$ if we assume w.l.o.g.~that $A$ is the better restaurant) up to the $n$th agent and $\phi_{k^*}$ is given in Proposition \ref{thm_infprob}. If we condition on $n^*<N$, then $A_{n^*}={(n^*+k^*)}/{2}$ with probability $\phi_{k^*}$ and ${(n^*-k^*)}/{2}$ with probability $1-\phi_{k^*}$. Thus, welfare is 
$$W_N(\bar \rho_N, w) = \mathbb{E}\bigg[\sum_{n=1}^\infty r_{n}  \mathbf{1}_{\{n^*=n\}}\bigg]$$
where
$$r_n = \begin{cases} \phi_{k^*} - \big(n-k^*\big) \big(2\phi_{k^*}-1\big)/2N & \text{if $n \le N$} \\  A_N/N & \text{if $n \ge N$} \end{cases}$$
Note that in writing $W_N$ as we have above, we are treating the finite agent model as a truncation of the infinite agent model, and that the $n=N$ case in the definition of $r_n$ above admits both expressions, a fact we will make use of below.  \bigskip

\noindent (i) Let us first prove that $W_N$ increases in $N$. Observe
\begin{align*}
W_N(\bar \rho_N, w) &= \mathbb{E}\bigg[\bigg[\phi_{k^*} - \frac{n^*-k^*}{2N}    \big(2\phi_{k^*}-1\big) \bigg]  \mathbf{1}_{\{n^*\le N\}} +\frac{A_N}{N}  \mathbf{1}_{\{n^*>N\}}\bigg]\\
&<\mathbb{E}\bigg[\bigg[\phi_{k^*} - \frac{n^*-k^*}{2(N+1)}    \big(2\phi_{k^*}-1\big) \bigg]  \mathbf{1}_{\{n^*< N+1\}} + \frac{A_N}{N}  \mathbf{1}_{\{n^*\ge N+1\}}\bigg].
\end{align*}
If
$$\mathbb{E}\bigg[\frac{A_N}{N}  \mathbf{1}_{\{n^*\ge N+1\}}\bigg] \le \mathbb{E}\bigg[\frac{A_{N+1}}{N+1}  \mathbf{1}_{\{n^*\ge N+1\}}\bigg],$$
then we can upper bound the second line with $W_{N+1}(\bar{\rho}_{N+1},\,w).$ Note that \[\mathbb{E}\big[A_{N+1}  \mathbf{1}_{\{n^*\ge N+1\}}\big] = \mathbb{E}\big[(A_{N}+p)  \mathbf{1}_{\{n^*\ge N+1\}}\big]\] since the difference solely depends on the private signal of the $(N+1)$th agent. Using this equation, the above display reduces to $\mathbb{E}\big[A_{N}  \mathbf{1}_{\{n^*>N\}}\big] \le \mathbb{E}\big[Np  \mathbf{1}_{\{n^*> N\}}\big]$, or equivalently, $\mathbb{E}\big[A_{N} \,\big| \, n^*> N\big] \le Np$. By induction, it suffices to show that 
\begin{align}
\forall n, \qquad \mathbb{E}\big[A_{n+1} \,\big| \, n^*> n+1\big] \le \mathbb{E}\big[A_{n} \,\big| \, n^*> n\big] +p.
\label{eq_claim_popsize}
\end{align}  

\vspace{5pt}
\noindent \textit{Case 1}--- $n$ and $k^*$ have different parities: $n^*>n+1$ is equivalent to $n^*> n$. Therefore, inequality \eqref{eq_claim_popsize} holds with equality from $\mathbb{E}\big[A_{n+1} - A_n \,\big|\,K_n\big]=p.$

\vspace{5pt}
\noindent \textit{Case 2}--- $n$ and $k^*$ have the same parity: $\mathbb{E}\big[A_{n+1} - A_n \,\big|\,K_n\big]=p$ when $\lvert K_n \rvert < k^*-1$ following the same logic as Case 1, so $\mathbb{E}\big[A_{n+1} \,\big| \, n^*> n+1\big] = \mathbb{E}\big[A_{n} \,\big| \, n^*> n,\,\lvert K_n \rvert < k^*-1\big] +p$. It suffices to prove $\mathbb{E}\big[A_{n} \,\big| \, n^*> n,\,\lvert K_n \rvert < k^*-1\big] \le \mathbb{E}\big[A_{n} \,\big| \, n^*> n,\,\lvert K_n \rvert = k^*-1\big]$. Since $A_n = (n + K_n)/2$ conditional on $n^* > n$ and $\Pr\big[K_n = k \,\big|\,\lvert K_n \rvert = k \big]=\phi_k$ by symmetry,
\begin{align*}
\mathbb{E}\big[A_{n} \,\big| \, n^*> n,\,\lvert K_n \rvert =k\big] = \frac{n}{2}+ \frac{(2\phi_k - 1)k}{2}
\end{align*}
given that $n$ has has the same parity with $k$. The conclusion follows as this conditional expectation increases in $k$. \\

For convergence, we first prove that $n^*$ has finite first moment.
\begin{lemma}
    \label{lemma_fullexp}
    $\mathbb{E}[n^*]<\infty.$
\end{lemma}
\begin{proof}
Note that for $t \geq 1$, $$\textstyle{\Pr \big[n^* > t \big] \le \Pr \big[-k^* < K_t < k^*\big] \le \Pr\big[\mathrm{Binom}(t,\,p) < \frac{t}{2}+k^*\big]}$$ since the left hand side requires $-k^* < K_n < k^*$ for all $n=1,\dots,t$ while the right hand side requires $K_n < k^*$ only for $n=t$. The right hand side is upper bounded again by $\Pr \big[\mathrm{Binom}(t,\,p) < (\frac{1}{2}+\varepsilon)t\big]$ for a small $\varepsilon>0$ and large enough $t$. By the result of \citet{ferrante2021bounds} in Proposition \ref{thm_cjtrate}, this probability is $O(\frac{1}{\sqrt{t}}e^{-ct})$ for some constant $c>0$, which shows $$\mathbb{E}[n^*] = \int_0^\infty \Pr [n^* > t ] dt \le \int_0^1 1 dt + O \bigg(\int_1^\infty \frac{1}{\sqrt{n}}e^{-cn} dt\bigg) <\infty$$ where the first equation is due to $n^* \ge 0$ and Tonelli's theorem.
\end{proof}

Given the result above, observe 
\begin{align*}
W_N(\bar \rho_N, w) = \bigg[\phi_{k^*} - \frac{\mathbb{E}[n^*\,|\,n^*\le N]-k^*}{2N}    \big(2\phi_{k^*}-1\big) \bigg]  \Pr \big[n^*\le N\big] &   \\
+\mathbb{E}\bigg[\frac{A_N}{N} \,\bigg|\,n^*> N \bigg]  \Pr\big[n^*> N\big] &. 
\end{align*}
We have $\lim_{N \rightarrow \infty} \Pr \big[n^*> N\big] = 0$ since the process stops in finite time almost surely, as shown in Proposition \ref{thm_infprob}. Moreover,  $\lim_{N \rightarrow \infty} \frac{1}{2N}{\mathbb{E}[n^*\,|\,n^*\le N]-k^*} = 0$ since the stopping time has a finite first moment by Lemma \ref{lemma_fullexp}. Thus, $\lim_{N \rightarrow \infty}  W_N(\bar \rho_N, w)= \phi_{k^*}=W(\bar{\rho},\,w)$. \bigskip

\noindent(ii) Three claims are due: (a) $\sup w^*(\bar \rho_N, N) < \frac{1}{2}$, (b) $\sup w^*(\bar \rho_N, N)$ does not decrease in $N$, and (c) $\lim_N\,\inf w^*(\bar \rho_N, N) = \tfrac12$.

\vspace{5pt}
\noindent (a) $W_N(\bar \rho_N, w)= p$ if $\sup w^*(\bar\rho_N,\,N)>1/2$, and
$$W_N(\bar \rho_N,\,0) \ge W_4(\bar \rho_4,\,0) = \frac{4p^2 + 3\cdot 2p^3(1-p) + 2\cdot 4p^2(1-p)^2 + 2p(1-p)^3}{4} > p.$$
Therefore, the maximum of $W_N(\bar\rho_N,\,\cdot)$ is achieved for some $w<1/2$. \vspace{5pt}

\noindent (b) 
For $w<\tilde{w}$, let $n^*$ and $\tilde{n}^*$ be random variables that denote the stopping time associated with $w$ and $\tilde{w}$. Rewrite welfare as
$$N \cdot W_N(\bar \rho_N, w) = \mathbb{E}\big[ \big[A_{n^*} +(N-n^*)  \phi_{k^*(w)}\big]\mathbf{1}_{\{n^* \le N\}} + A_N  \mathbf{1}_{\{n^*> N\}}\big]$$
and
$$N \cdot W_N(\bar \rho_N, \tilde{w})= \mathbb{E}\big[ \big[A_{\tilde{n}^*} +(N-\tilde{n}^*)  \phi_{k^*(\tilde{w})}\big]\mathbf{1}_{\{\tilde{n}^* \le N\}} + A_{N} \mathbf{1}_{\{\tilde{n}^* > N\}}\big].$$
Since $\{n^* \le N\} \supset \{\tilde{n}^* \le N\}$,
\begin{align*}
    &N \cdot \big[W_N(\bar \rho_N, \tilde{w})-W_N(\bar \rho_N, w)\big]\\
    = &\mathbb{E} \big[\big[\big[A_{\tilde{n}^*}  + (N-\tilde{n}^* ) \phi_{k^*(\tilde{w})} \big] \mathbf{1}_{\{\tilde{n}^* \le N\}}  +A_{N}   \mathbf{1}_{\{\tilde{n}^* > N\}} \\
    & \hspace{6cm}-A_{n^*}-(N-n^*)  \phi_{k^*(w)} \big] \mathbf{1}_{\{n^* \le N\}}\big]\\
    = &\mathbb{E} \big[\big[\big[(A_{\tilde{n}^*}-A_{n^*})  + (N-\tilde{n}^* ) \phi_{k^*(\tilde{w})} \big] \mathbf{1}_{\{\tilde{n}^* \le N\}}  +(A_{N}-A_{n^*})   \mathbf{1}_{\{\tilde{n}^* > N\}} \\
    & \hspace{6cm}-(N-n^*)  \phi_{k^*(w)} \big] \mathbf{1}_{\{n^* \le N\}}\big].
\end{align*}
Assume the following inequalities: 
\begin{itemize}
\item[(1)] $\mathbb{E}\big[(A_{\tilde{n}^*}-A_{n^*}) \mathbf{1}_{\{\tilde{n}^* \le N\}}\big] \le \mathbb{E}\big[\phi_{k^*(\tilde{w})}(\tilde{n}^* - n^*)  \mathbf{1}_{\{\tilde{n}^* \le N\}}\big]$, \item[(2)] $\mathbb{E}\big[(A_{N}-A_{n^*})   \mathbf{1}_{\{\tilde{n}^* > N,\,n^* \le N\}} \big] \le \mathbb{E}\big[p(N-n^*)   \mathbf{1}_{\{\tilde{n}^* > N,\,n^* \le N\}} \big]$.
\end{itemize} If we plug these in the above expectation, then we obtain that $W_N(\bar \rho_N, \tilde{w}) \ge W_N(\bar \rho_N, w)$ implies 
$$\mathbb{E} \big[\big[\phi_{k^*(\tilde{w})} \mathbf{1}_{\{\tilde{n}^* \le N\}}  +p   \mathbf{1}_{\{\tilde{n}^* > N\}} - \phi_{k^*(w)} \big] \mathbf{1}_{\{n^* \le N\}}\big] \ge 0,$$
which is a rearrangement of 
\begin{align*}
&\big[(N+1) \cdot W_{N+1}(\bar \rho_{N+1},\,\tilde{w}) - N \cdot W_N(\bar \rho_N, \tilde{w})\big] \\
& \hspace{3cm} - \big[(N+1)\cdot W_{N+1}(\bar \rho_{N+1},\,w) - N\cdot W_N(\bar \rho_N, w)\big]\\
=&\mathbb{E}\big[\phi_{k^*(\tilde{w})}  \mathbf{1}_{\{\tilde{n}^* \le N\}} + p  \mathbf{1}_{\{\tilde{n}^* > N\}}\big] - \mathbb{E}\big[\phi_{k^*(w)}  \mathbf{1}_{\{n^* \le N\}} + p  \mathbf{1}_{\{n^* > N\}}\big].
\end{align*}
This is to say if $W_N(\bar \rho_N, \tilde{w}) \ge W_N(\bar \rho_N, w)$, then $W_{N+1}(\bar \rho_{N+1}, \tilde{w}) \ge W_{N+1}(\bar \rho_{N+1}, w)$. It follows that $\sup w^*(\bar\rho_N,\,N) \le \sup w^*(\bar\rho_{N+1},\,N+1)$.

We finally justify (1) and (2) above. For inequality (1), due to symmetry, any sequence of signals that ends at $K_{\tilde{n}^*}=k^*(\tilde{w})$ is $\phi_{k^*(\tilde{w})}$ times more likely than the symmetric sequence that ends at $K_{\tilde{n}^*}=-k^*(\tilde{w})$. For any choice of $\{\rho_n\}_{n=1}^N$ and its completely symmetric restaurant choices $\{\neg \rho_n\}_{n=1}^N$, the following holds for $\delta =  \mathbb{E}\big[A_{\tilde{n}^*}-A_{n^*}\,|\, \{\rho_n\}_{n=1}^N ]$: 
\begin{align*}
    &\mathbb{E}\big[A_{\tilde{n}^*}-A_{n^*} \,|\, \{\rho_n\}_{n=1}^N \cup \{\neg\,\rho_n\}_{n=1}^N\big] \\
    &= \mathbb{E}\big[A_{\tilde{n}^*}-A_{n^*}\,|\, \{\rho_n\}_{n=1}^N ]  \Pr\big[\{\rho_n\}_{n=1}^N \,|\, \{\rho_n\}_{n=1}^N \cup \{\neg\,\rho_n\}_{n=1}^N\big] \\
    & \hspace{12pt} +\mathbb{E}\big[A_{\tilde{n}^*}-A_{n^*}\,|\, \{\neg \rho_n\}_{n=1}^N ]  \Pr\big[\{\neg\,\rho_n\}_{n=1}^N\,|\, \{\rho_n\}_{n=1}^N \cup \{\neg\,\rho_n\}_{n=1}^N\big]  \\
    & = \phi_{k^*(\tilde{w})}  \delta + (1-\phi_{k^*(\tilde{w})}) (\tilde{n}^* -n^* - \delta) \\
    &\le \phi_{k^*(\tilde{w})}(\tilde{n}^*-n^*).
\end{align*}
Taking the expectation over $\{\{\rho_n\}_{n=1}^N :\tilde{n}^* \le N\}$ proves the claim. 

For inequality (2), we can show that fixing $n^* = t  \le N$, we have $\mathbb{E}\big[A_{n+1}-A_{n^*} \,\big| \, \tilde{n}^*>n+1,\,n^* =t \big] \le \mathbb{E}\big[A_{n} -A_{n^*}\,\big| \, \tilde{n}^*> n,\,n^* = t\big] +p$ for all $n \ge t$, by repeating the proof of inequality \eqref{eq_claim_popsize} in part (i). Taking the expectation over $\{t \le N\}$ proves the claim. \vspace{5pt} 

\noindent (c) Fix $k^*(\tilde w)$ for some $\tilde w$. Then  $\lim_{N\rightarrow \infty} W_N(\bar \rho_N, w) = \phi_{k^*(w)}$ by part (i). Convergence occurs only through the value of $k^*(w)$, so $\lim_{N\rightarrow \infty} \sup_{w\,:\,k^*(w) \le k^*(\tilde w)} \lvert W_N(\bar \rho_N, w) - \phi_{k^*(w)} \rvert = 0$. Since $\phi_{k^*(w)}$ strongly increases in $k^*(w)$, there is $N_{k^*(\tilde w)}$ such that for any $N \ge N_{k^*(\tilde w)}$, \[W_N(\bar \rho_N, w) \approx \phi_{k^*(w)} <  \phi_{k^*(\tilde w)}  \approx W_N(\bar \rho_N,\,\tilde w) \qquad \forall w, \tilde w \] where $k^* (w) < k^*(\tilde w)$. Therefore, the maximum cannot be attained when $k^*(w) < k^*(\tilde w)$ for large $N$. The claim follows as $k^*(\tilde w)$ can be arbitrarily large. 

\subsection{Proof of Proposition \ref{thm_social_rate}}


To derive the convergence rates, we  need to understand the asymptotic behavior of $n^*$. \citet[p.~350]{feller1950introduction} characterizes the distribution of $n^*$ using generating functions, but we derive its full probability mass function using the transition matrix. \bigskip

\noindent (i) Assume even $k^*(w)$. Let $p_{n,\,k}$ ber $\Pr[K_n = k,\,n^*\ge n]$, the probability that $K_n$ has not hit the barriers until period $n$ and $K_n=k$, and $p_n =(p_{n,\,-k^*(w)}, p_{n,\,-k^*(w)+2},\dots,p_{n,\,k^*(w)})^\top \in \mathbb{R}^{k^*(w) +1}$. By definition, $p_0$ puts a unit mass on $p_{0,0}$. The process follows
$$p_{2n+2} = 
\begin{pmatrix} 
0 & p^2 & 0 & 0 & 0 &  \cdots \\
0 & 2p(1-p) & p^2 & 0 & 0 & \cdots \\
0 & (1-p)^2 & 2p(1-p) & p^2 & 0  & \cdots \\
0 & 0 & (1-p)^2 & 2p(1-p) & p^2 & \cdots \\
\vdots
\end{pmatrix} 
\times p_{2n}.$$
Note that $\Pr[n^*=n] = p_{n,\,k^*(w)}+p_{n,\,-k^*(w)}$, the probability that a cascade happens at agent $n$. Denote the transition matrix as $T$. The analysis simplifies if $T$ is diagonalizable. In fact, $T=V\Lambda V^{-1}$ holds for the following $V\in \mathbb{R}^{(k^*(w)+1)\times(k^*(w)+1)}$ and $\Lambda \in \mathrm{diag}(\mathbb{R}^{k^*(w)+1})$:
$$\Lambda = 4p(1-p) \cdot \mathrm{diag}\bigg(\cos^2 \frac{\pi}{2k^*(w)},\,\cos^2 \frac{2\pi}{2k^*(w)},\,\cdots,\, \cos^2 \frac{(k^*(w)-1)\pi}{2k^*(w)},\,0,\,0 \bigg)$$
and
$$V = 
\begin{pmatrix} \mathbf{1}_{k^*(w)-1}^\top & 1 & 0 \\
M & \mathbf{0}_{k^*(w)-1} & \mathbf{0}_{k^*(w)-1} \\
v^\top & 0 & 1 \\
\end{pmatrix},\,\,\,V^{-1} = 
\begin{pmatrix} 
\mathbf{0}_{k^*(w)-1} & M^{-1} & \mathbf{0}_{k^*(w)-1} \\
1 & -\mathbf{1}_{k^*(w)-1}^\top M^{-1} & 0 \\
0 & -v^\top M^{-1} & 1 \\
\end{pmatrix}$$
for matrix $M\in \mathbb{R}^{(k^*(w)-1)\times (k^*(w)-1)}$ and its inverse $M^{-1}$ whose $(i,j)$th entries are
$$M_{ij} = 2\cot \frac{\pi j}{2k^*(w)}\sin \frac{\pi ij}{k^*(w)}  \bigg(\frac{1-p}{p}\bigg)^{i},\,\,\,(M^{-1})_{ij} =  \frac{1}{2} \tan \frac{\pi i}{2k^*(w)} \sin \frac{\pi ij}{k^*(w)}  \bigg(\frac{p}{1-p}\bigg)^{j}$$
and vector $v\in \mathbb{R}^{k^*(w)-1}$ whose $i$th entry is 
$$v_i = \bigg(\frac{1-p}{p}\bigg)^{k^*(w)}\cdot (-1)^{i+1}.$$
The proof is via a straightforward calculation using the fact that $$\sum_{r=1}^{k^*(w)-1} \sin \frac{\pi ir}{k^*(w)}\sin \frac{\pi jr}{k^*(w)} = \frac{k^*(w)-1}{2}  \delta_{ij}$$
where $\delta_{ij}$ is the Kronecker delta. Note that $p_{2n} = V \Lambda^{n} V^{-1} p_0$ by diagonalization, so
\begin{align*}
\Pr[n^*\ge 2N] & = (e_1 + e_{k^*(w)+1})^\top V \bigg[ \sum_{n=N}^{\infty} \Lambda^n \bigg] V^{-1} p_0 \\
& = (e_1 + e_{k^*(w)+1})^\top V \mathrm{diag}\bigg(\frac{\lambda_i^N}{1-\lambda_i}\bigg) V^{-1} p_0
\end{align*}
where $e_1$ and $e_{k^*(w)+1}$ extract the first and last elements of the ensuing vector, and $\lambda_i = \Lambda_{ii}$. Plugging in the explicit form of the largest eigenvalue proves the claim. For odd $k^*(w)$, the construction is identical except for using $p_1$ as the initial value instead of $p_0$. \bigskip

\noindent (ii) Note that $\phi_{k^*(w)} - W_N(\bar \rho_N, w)$ is 
$$\big(2\phi_{k^*(w)} -1\big) \sum_{n=1}^N \frac{(n-k^*(w))  \mathbf{1}_{\{n^* = n\}}}{2N} + \mathbb{E}\bigg[\bigg(\phi_{k^*(w)} - \frac{A_N}{N}\bigg) \mathbf{1}_{\{n^* > N\}}\bigg].$$
The first term is $\Theta(\mathbb{E}[n^*\cdot  \mathbf{1}_{\{n^* \le N\}}]/N)$, and the second term is $O(\Pr[n^*> N]).$ Note that $\Pr[n^*=k] < \mathbb{E}[n^*  \cdot \mathbf{1}_{\{n^* \le N\}}] < \mathbb{E}[n^*]$. The lower bound is $\Theta(1)$ as it does not depend on $N$, and the upper bound is $\Theta(1)$ by Lemma \ref{lemma_fullexp}. Meanwhile, we have derived above that $\Pr[n^* > N]= O(\lambda_1^{N/2})=o(1/N)$ for the largest eigenvalue $\lambda_1$ associated with the threshold $k^*(w)$. The overall expression reduces to $\Theta(1/N)$, completing the proof.

\clearpage

\bibliography{refs}

\end{document}